\documentstyle[12pt,epsf]{article}
\advance\voffset by -1.5cm

\newcommand{\be}{\begin{eqnarray}}
\newcommand{\ee}{\end{eqnarray}}

\title{\begin{flushright}
{\normalsize McGill/98-14\\NUC-MINN-98/7-T\\
August 1998 \\}
\end{flushright}
\vspace*{0.1in}
{\bf COHERENCE TIME IN HIGH ENERGY PROTON-NUCLEUS COLLISIONS}}
\author{{\bf Charles Gale}$^1$ \vspace*{0.1in} \\
 {\it Physics Department}\\
 {\it McGill University}\\ \vspace*{0.2in}
 {\it Montreal, Quebec H3A 2T8, Canada}\\ \vspace*{0.1in}
{\bf Sangyong Jeon}$^2$ and {\bf Joseph Kapusta}$^3$ \\
  {\it School of Physics and Astronomy}\\
  {\it University of Minnesota}\\ {\it Minneapolis, MN 55455}}

\date{}

\begin{document}

\maketitle
\begin{abstract}

Precisely measured Drell-Yan cross sections for 800 GeV protons incident
on a variety of nuclear targets exhibit a deviation from linear scaling in the
atomic number $A$.  We show that this deviation can be accounted for by
energy degradation of the proton as it passes through the nucleus if
account is taken of the time delay of particle production due to
quantum coherence.  We infer an average proper coherence time of
0.4$\pm$0.1 fm/c, corresponding to a coherence path length of
8$\pm$2 fm in the rest frame of the nucleus.

\end{abstract}



\noindent
$^1$ gale@hep.physics.mcgill.ca\\
$^2$ jeon@nucth1.spa.umn.edu\\
$^3$ kapusta@physics.spa.umn.edu\\

\newpage

Propagation of a high energy particle through a medium is of interest
in many areas of physics.  High energy proton-nucleus scattering has
been studied for many decades by both the nuclear and particle physics
communities \cite{Wit}.  Such studies are particularly relevant for
the Relativistic Heavy Ion Collider (RHIC), which will collide beams
of gold nuclei at an energy of 100 GeV per nucleon, and for the Large
Hadron Collider (LHC), which will collide beams of lead nuclei at
1500 GeV per nucleon \cite{qm96}.

There are two extreme limits of a projectile scattering from a nucleus.
When the cross section of the projectile with a nucleon is very small,
as is the case for neutrinos, Glauber theory says that the cross section
with a nuclear target of atomic number $A$ grows linearly with $A$.  When
the cross section with an individual nucleon is very large, as is
the case for pions near the delta resonance peak, the nucleus appears
black and the cross section grows like $A^{2/3}$.  A more interesting
case is the production of lepton pairs with large invariant mass,
often referred to as Drell-Yan, in proton-nucleus
collisions.  Both the elastic and inelastic cross sections for
proton-nucleon scattering are relatively large, but the partial
cross section to produce a high mass lepton pair, being electromagnetic
in origin, is relatively small.  Experiments have shown that the
inclusive Drell-Yan cross section grows with $A$ to a power very
close to 1.  The theoretical interpretation is that the hard
particles, the high invariant mass lepton pairs, appear first
and the soft particles, the typical mesons, appear later
due to quantum-mechanical interference, essentially the uncertainty
principle.  These quantum coherence requirements also lead to the
Landau-Pomeranchuk-Migdal effect \cite{lpm}.  Deviations from the
power 1 by high precision Drell-Yan experiments \cite{E772}
at Fermi National Accelerator Laboratory (FNAL) suggest
that it may be possible to infer a finite numerical value for the
coherence time.  That is the goal of this paper.

For a basic description of high energy proton-nucleus scattering
we make a straightforward linear extrapolation from proton-proton
scattering.  This extrapolation, referred to as LEXUS, was detailed
and applied to nucleus-nucleus collisions at beam energies of several
hundred GeV per nucleon in ref. \cite{lexus}.  Briefly, the
inclusive distribution in rapidity $y$ of the beam proton in an
elementary proton-nucleon collision is parameterized rather well by
\begin{equation}
W_1(y) = \lambda \frac{\cosh y}{\sinh y_0}
+ (1-\lambda)\delta(y_0-y) \, ,
\end{equation}
where $y_0$ is the beam rapidity in the lab frame.
The parameter $\lambda$ has the value 0.6 independent of beam
energy, at least in the range in which it has been measured, which
is $12-400$ GeV.  It may be interpreted as the fraction of all
collisions which are neither diffractive nor elastic.
As the proton cascades through the nucleus its
energy is degraded.  Its rapidity distribution satisfies an
evolution equation \cite{hwa84} whose solution is, after $i$ collisions
\cite{ck85}:
\begin{equation}
W_i(y) = \frac{\cosh y}{\sinh y_0} \sum_{k=1}^i
\left( \begin{array}{c} i \\ k \end{array} \right)
\frac{\lambda^k (1-\lambda)^{i-k}}{(k-1)!}
\left[ \ln\left( \frac{\sinh y_0}{\sinh y} \right) \right]^{k-1}
+ (1-\lambda)^i \delta(y_0-y) \, .
\end{equation}
This distribution then gets folded with impact parameter over the
density distribution of the target nucleus as measured by electron
scattering.  Every inelastic collision produces an average number
of negatively charged hadrons given by the simple formula
\begin{equation}
\langle h^- \rangle_{NN} = 0.784 \,
 \frac{(\sqrt{s} - 2m_N -m_{\pi})^{3/4}}{s^{1/8}} \, .
\end{equation}
The negatively charged hadrons are Gaussian distributed in rapidity with
a width given by $\sqrt{\ln(\sqrt{s}/2m_N)}$ in customary notation.

The rapidity distribution of negatively charged hadrons, as computed
in the way described, is nearly centered in the nucleon-nucleon
center-of-momentum (c.m.) frame, whereas data taken for p+S and p+Au
collisions at 200 GeV \cite{NA35} are skewed towards the target rest frame.
If one allows for a small rapidity shift of 0.16 whenever a produced
hadron encounters a struck target nucleon, chosen with a sign corresponding
to a slowing down of the hadron relative to the nucleon, one obtains
the curves shown in Fig. 1.  The paucity of computed hadrons compared
to data at small rapidity in p+Au collisions is undoubtedly due to a
further cascading and particle production by struck nucleons in such a
large nucleus.  This physics could be incorporated by a more detailed
cascade code followed by nuclear evaporation, but is not essential
for our purposes in this paper.  Apart from that, the description
of the data is very good, including absolute normalization,
especially considering that there is only one free parameter.

As the proton cascades through the nucleus it undergoes a random walk
in transverse velocity.  This broadens the transverse momentum
distribution of the produced hadrons relative to pp collisions in the
way described in ref. \cite{lexus}.  The transverse momentum
distributions, for various windows of rapidity, are shown in
Fig. 2.  There are no free parameters apart from the rapidity
shift which was already fitted to be 0.16.

Now we turn to a description of the Drell-Yan.  Figure 3 is a schematic
of two limits and an intermediate situation.  One limit is full
energy degradation of the proton as it traverses the nucleus.  Produced
hadrons appear immediately with zero coherence time, causing the proton
to have less energy available to produce a Drell-Yan pair at the backside
of the nucleus.  The other limit is usually referred to as Glauber,
although this is a bit of a misnomer.  Produced hadrons, being soft
on the average, do not appear until after the hardest particles, the
Drell-Yan pair, have already appeared.  This is the limit of a very
large coherence time, and it allows the proton to produce the Drell-Yan
pair anywhere along its path with the full incident beam energy.
An intermediate case is one of finite, nonzero coherence time.
By the time the proton wants to make a Drell-Yan pair on the backside
of the nucleus, hadrons have already appeared from the first collision
but not from the second.  Therefore the proton has more energy available
to produce the Drell-Yan pair than full zero coherence time but less
energy than with infinite coherence time.  This ought to result
in an $A$ dependence less than 1, with the numerical value determined
by the coherence time.

To compute the Drell-Yan yields we use the parton model with the
GRV structure functions \cite{grv94} to leading order with a K factor.
These structure functions distinguish between pp and pn collisions.
We have compared the results to pp collisions at the same beam energy
of 800 GeV \cite{800pp} and
found the agreement to be excellent for all values of $x_F$.

The experiment E772 \cite{E772} measured the ratio
$\sigma^{\rm DY}_{pA}/(\sigma^{\rm DY}_{pd}/2)$.  Were there no energy
loss and all nuclei were charge symmetric this ratio would be
equal to $A$.  The experiment measured muon pairs with invariant mass
$M$ between 4 and 9 GeV and greater than 11 GeV to eliminate
the $J/\psi$ and $\Upsilon$ contributions.
The data has been presented in 7 bins of
Feynman $x_F$ from 0.05 to 0.65.  (Recall that $x_F$ is the ratio of
the muon pair longitudinal momentum to the incident beam momentum in the
nucleon-nucleon c.m. frame.)  Data for exemplary values of
0.05, 0.35 and 0.65 are shown in Fig. 4.  The data should fall on
the dashed line if the ratio of cross sections is $A$.  There is
a small but noticeable departure for tungsten and at the largest value
of $x_F$.  This is to be expected if energy loss plays a role as it
must affect the largest target nucleus and the highest energy muon
pairs the most.

We have computed the individual cross sections $\sigma^{\rm DY}_{pA}$
with a variable time delay.  The proton cascades through the nucleus
as described earlier, but we assume that the energy available to
produce a Drell-Yan pair is that which the proton has after $n$
previous collisions.  Thus $n = 1$ is full energy loss and $n = \infty$
is zero energy loss.  We have taken the resulting proton-nucleus
Drell-Yan cross section, multiplied it by 2, divided it by the
sum of the computed pp and pn cross sections and display the results
in Fig. 4.  The lower edge of the shaded regions in the figure
corresponds to $n = 4$ and the upper edge to $n = 6$.
Overall the best representation of the data lies somewhere in this
range.  This collision number shift is easily converted to a coherence
time.  Let $\tau_0$ be the coherence time in the c.m. frame of the
colliding nucleons.  This is essentially the same as the formation
time of a pion since most pions are produced with rapidities near
zero in that frame.  The first proton-nucleon collision is the most
important, so boosting this time into the rest frame of the target
nucleus and converting it to a path length (proton moves essentially
at the speed of light) gives $\gamma_{\rm cm}\,c\,\tau_0 \approx
\sqrt{\gamma_{\rm lab}/2}\,c\,\tau_0$.  This path length may then be
equated with $n$ times the mean free path
$l = 1/\sigma^{\rm tot}_{\rm NN} \rho$.  Using a total cross section
of 40 mb and a nuclear matter density of 0.155 nucleons/fm$^3$ we
obtain a path length of 8$\pm$2 fm and a proper coherence time
of 0.4$\pm$0.1 fm/c corresponding to $n=5\pm 1$.

The analysis performed here can and should be improved upon.  What
we have done is a rough approximation to adding the quantum mechanical
amplitudes for a proton scattering from individual nucleons within
a nucleus.  A more sophisticated treatment would undoubtedly lead
to even better agreement with experiment, but the inferred value of
the proper coherence time is unlikely to be much different than obtained
with this first estimate.  The implications for nucleus-nucleus collisions
\cite{kaha2} will undoubtedly be important; they are under investigation.

\section*{Acknowledgements}

C.G. thanks the School of Physics and Astronomy at the University of
Minnesota for its hospitality during a sabbatical leave.
C.G. and J.K. thank the Institute for Nuclear Theory at the University
of Washington for its hospitality and the Department of Energy for
partial support during the program "Probes of Dense Matter in
Ultrarelativistic Heavy Ion Collisions".  This work was also supported
by the U. S. Department of Energy under grant
DE-FG02-87ER40328, by the Natural Sciences and Engineering Research Council
of Canada, and by the Fonds FCAR of the Quebec Government.

\newpage

\section*{Figures}

\begin{figure}[htb]
\epsfxsize=\textwidth
\centerline{\epsfbox{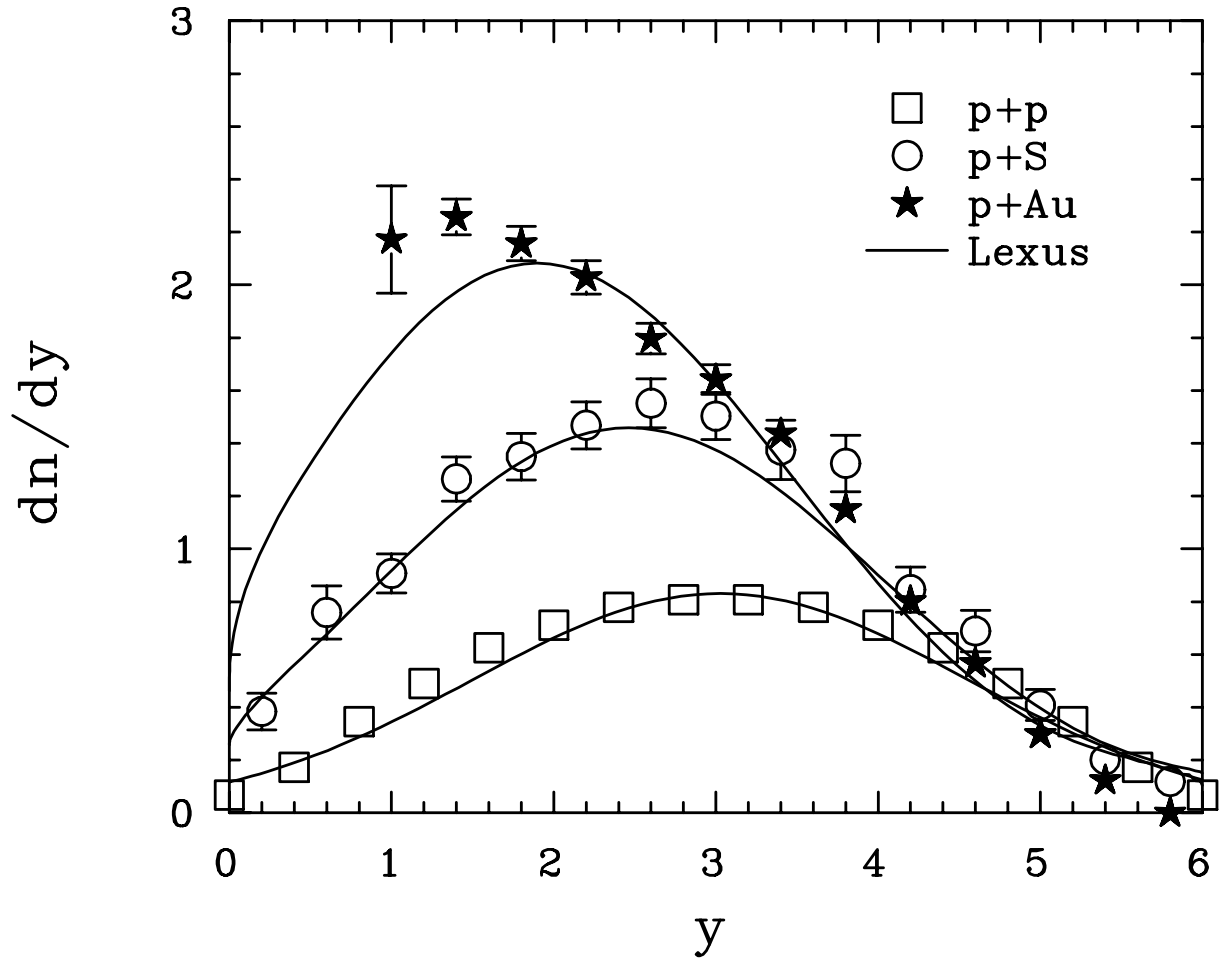}}
\caption{
The rapidity distribution of negatively charged hadrons
in pp, p+S, and p+Au collisions at a beam energy of 200 GeV.
The pp data are from \cite{pp} and the p+S and p+Au data are
from NA35 \cite{NA35}.  The curves are calculated with LEXUS with
a rapidity shift of 0.16 per struck nucleon.}
\end{figure}

\clearpage

\begin{figure}[htb]
\epsfysize=18cm
\centerline{\epsfbox{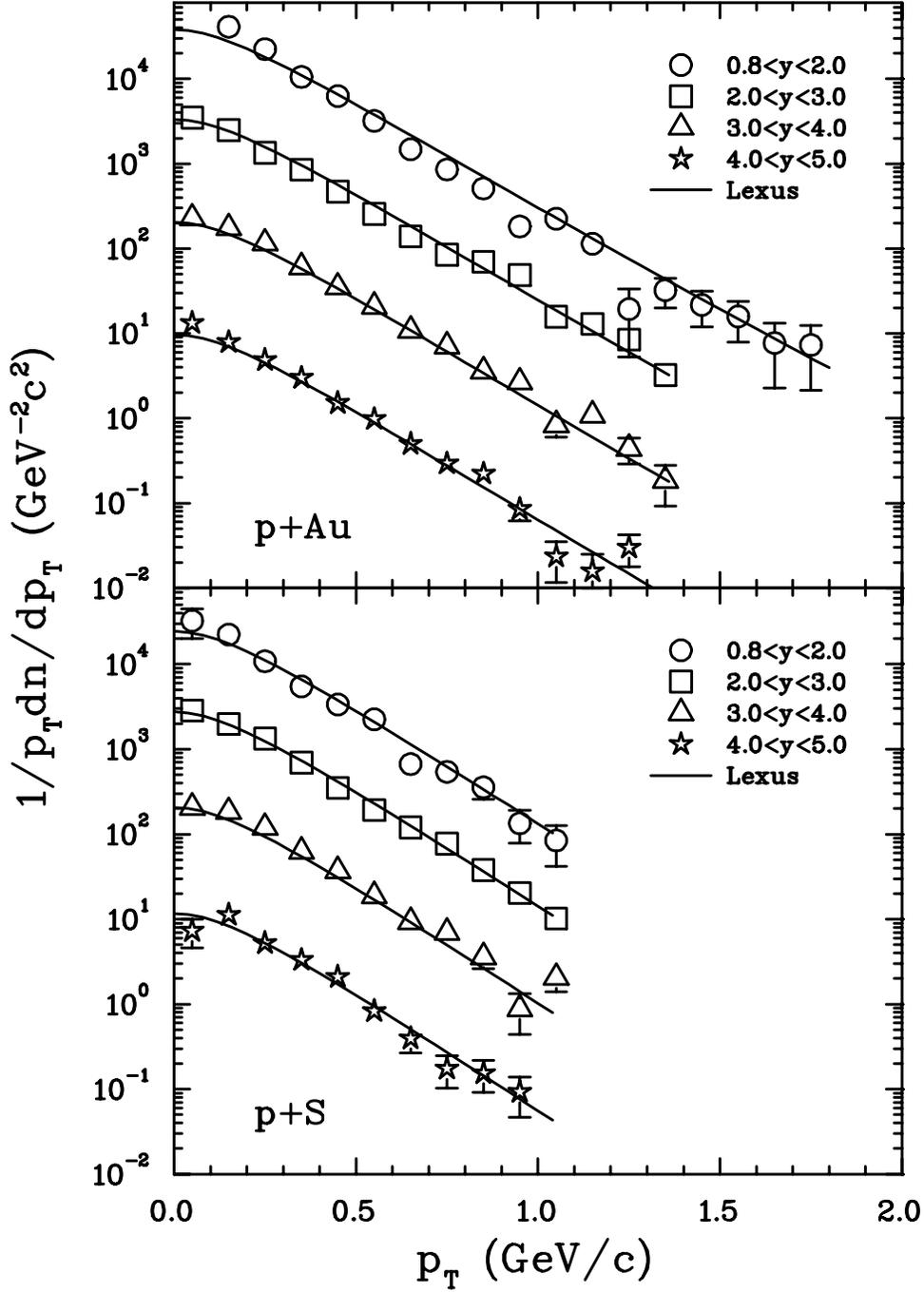}}
\caption{
The transverse momentum distributions for the same p+S
and p+Au collisions as in Fig. 1.  The curves are calculated with
LEXUS with a rapidity shift of 0.16 per struck nucleon.}
\end{figure}

\clearpage

\begin{figure}[htb]
\epsfysize=18cm
\centerline{\epsfbox{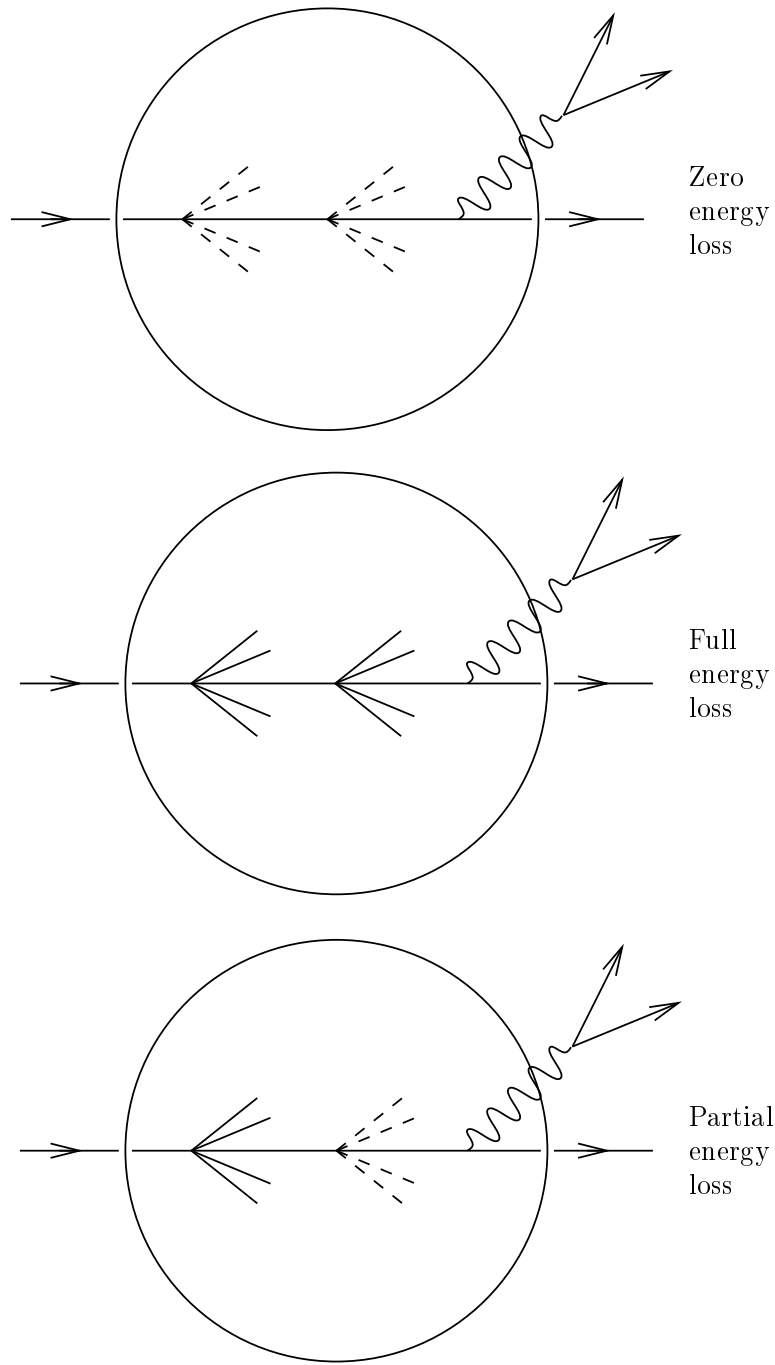}}
\caption{
Schematic of a high energy proton passing through a nucleus.
The upper panel represents no energy loss: the usual Glauber picture.
The middle panel represents full energy loss: hadrons are produced
immediately and the proton has less energy available for each subsequent
collision.  
The bottom panel represents partial energy loss: there is
a finite time delay before hadrons are produced and so the proton
has more energy available to create a high energy Drell-Yan pair.
}
\end{figure}

\clearpage

\begin{figure}[htb]
\epsfxsize=\textwidth
\centerline{\epsfbox{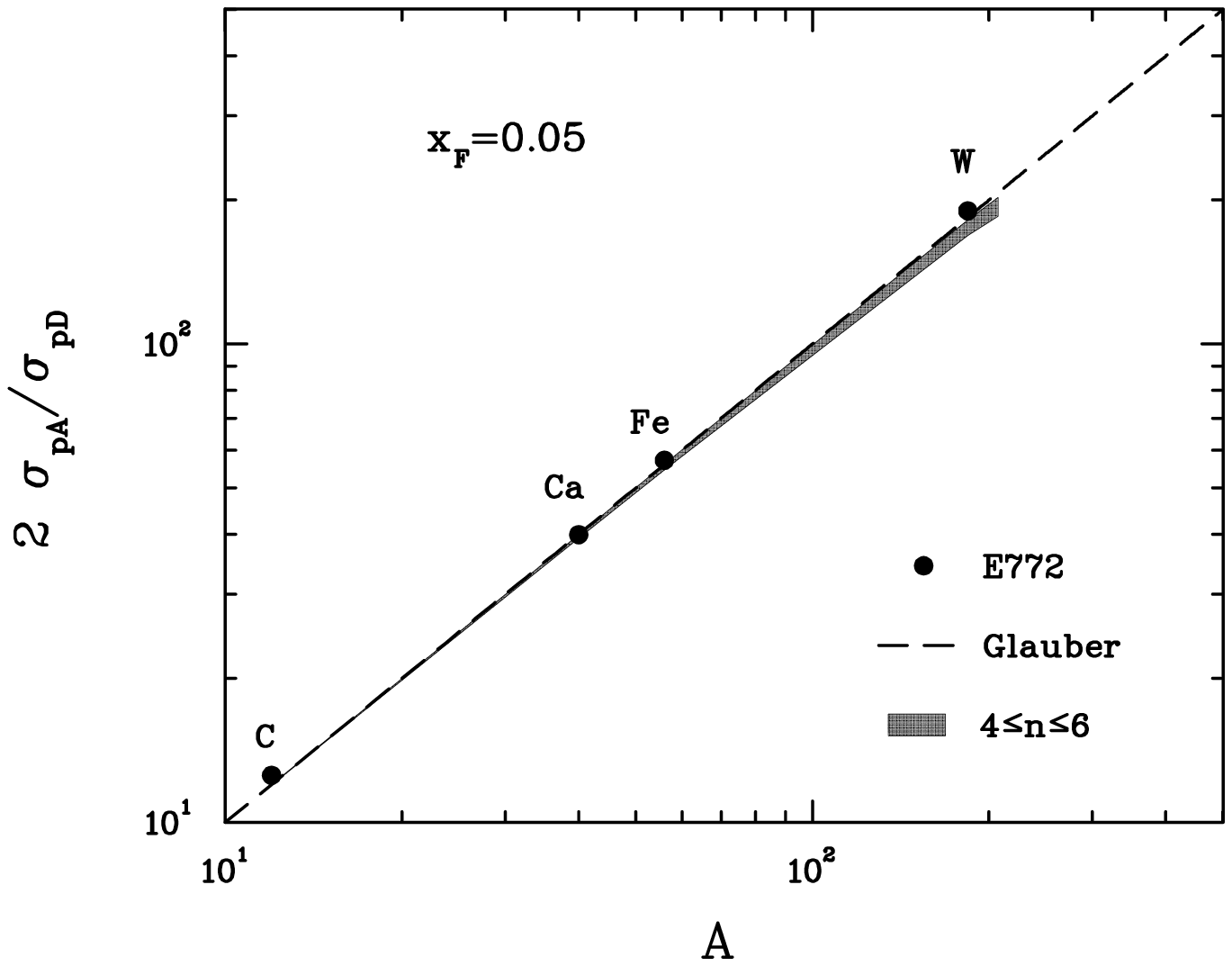}}
Figure 4-1: The ratio of the pA Drell-Yan cross section to the
proton-deuterium cross section divided by 2 for a beam energy
of 800 GeV.  The data are from E772 \cite{E772}.  The dashed
line assumes a scaling linear in the atomic number A.  The shaded
region represents our calculations with a coherence time ranging
from 4 to 6 proton-nucleon collisions, both elastic and inelastic.
We computed with target nuclei C, Ca, Fe, W and Pb and interpolate
between with straight lines to guide the eye.  The values of Feynman
$x_F$ of the Drell-Yan pair is indicated in each panel.
\end{figure}
\clearpage

\begin{figure}[htb]
\epsfxsize=\textwidth
\centerline{\epsfbox{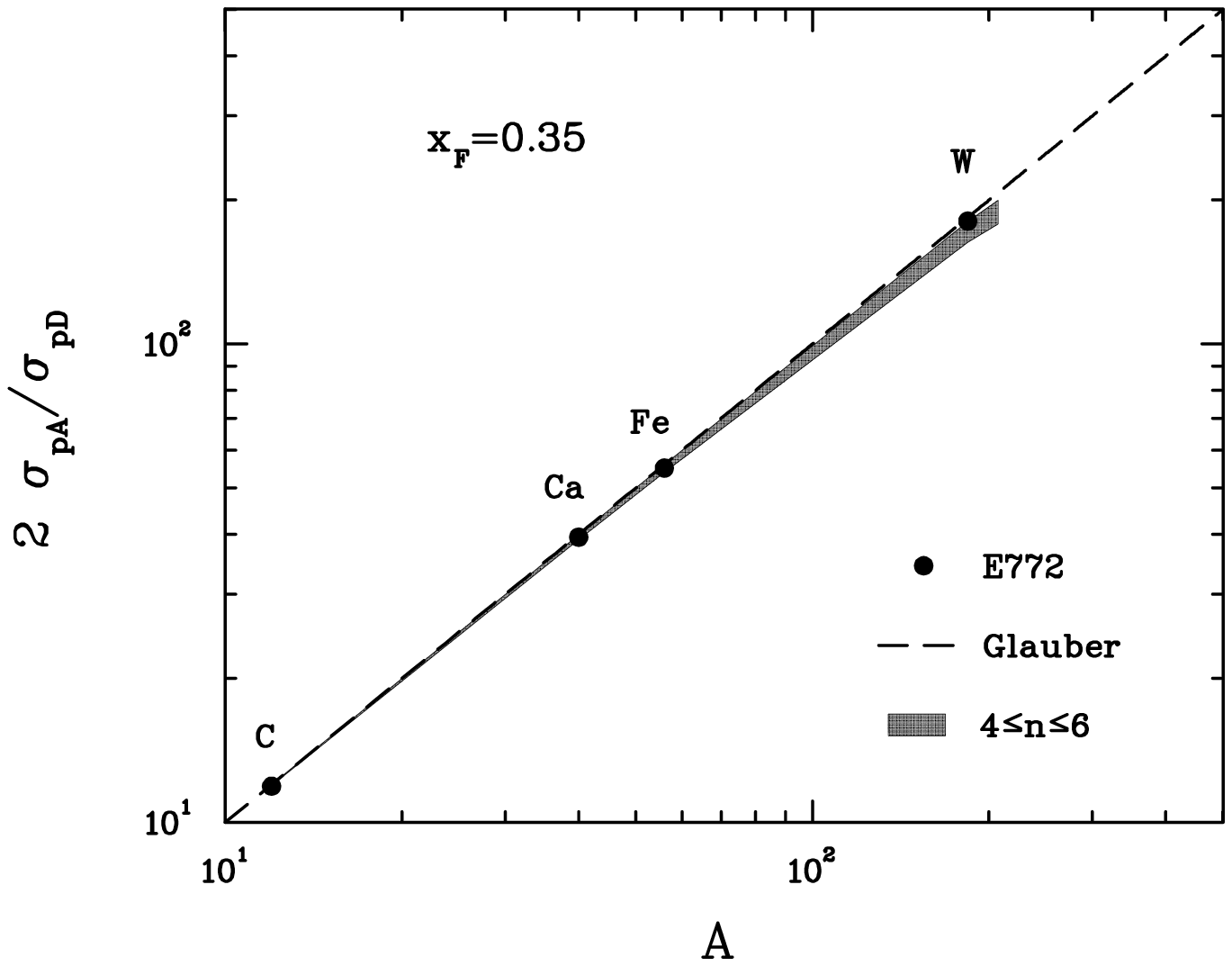}}
Figure 4-2: See the caption of Figure 4-1.
\end{figure}

\clearpage

\begin{figure}[htb]
\epsfxsize=\textwidth
\centerline{\epsfbox{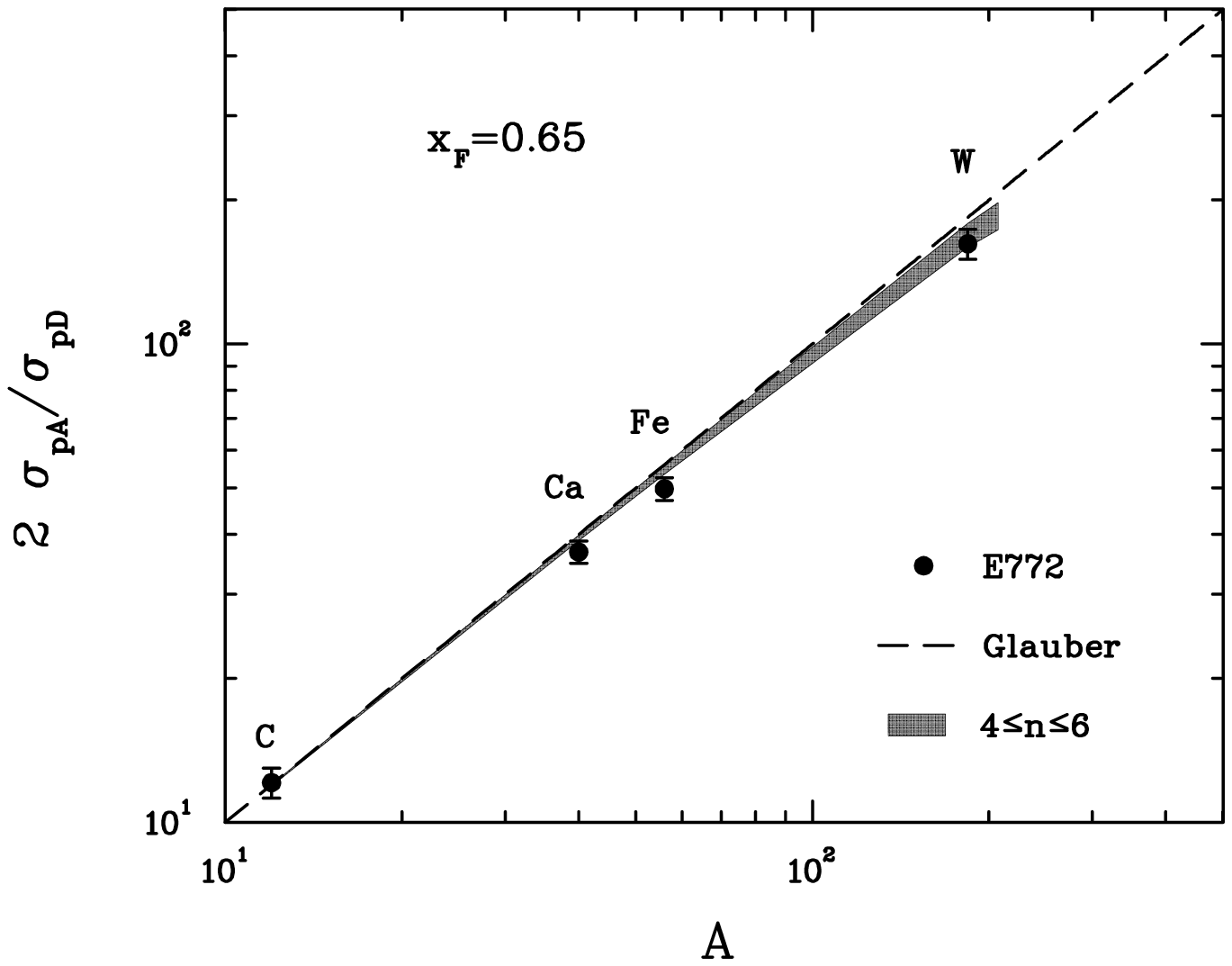}}
Figure 4-3: See the caption of Figure 4-1.
\end{figure}

\end{document}